\documentclass{elsart5p}
\usepackage{amssymb}
\usepackage{graphicx}  % standard LaTeX graphics tool
\usepackage{wasysym}
\usepackage{xcolor}
\usepackage{caption}
\usepackage{verbatim}
\usepackage{float}
\begin{document}
\begin{frontmatter}

\title{Status on the DESIR High Resolution Separator Commissioning}
\author[label1]{J. Michaud$^1$,}
\author[label1]{P. Alfaurt,} 
\author[label1]{A. Balana,} 
\author[label1]{B. Blank,} 
\author[label1]{L. Daudin,} 
\author[label1]{T. Kurtukian Nieto,} 
\author[label1]{B. Lachacinski,}
\author[label1]{L. Serani,} 
\author[label2]{F. Varenne,}

\address[label1]{Univ. Bordeaux, CNRS, LP2I Bordeaux, UMR 5797, F-33170 Gradignan, France}
\address[label2]{Grand Acc\'el\'erateur National d'Ions Lourds, 
        CEA/DRF - CNRS/IN2P3, Bvd Henri Becquerel, BP 55027, F-14076 CAEN Cedex 5, France}
        
\begin{abstract}

Many nuclear reactions used to create radioactive isotopes for nuclear research produce, in addition to the isotope of interest, many contaminants, which are often produced in much larger amounts than the isotope of interest. Many installations using the ISOL approach are therefore equipped with high-resolution mass separators to remove at least isotopes with a different mass number. In the present paper, we present the results of the commissioning of the DESIR HRS presently under development at LP2I Bordeaux (formerly CENBG). Optical aberrations are corrected up to 3rd order and a mass resolution of $M/\Delta M$ of 25000 is reached with a transmission of about 70\% for a $ ^{133}Cs^{+} $ beam at 25 keV.

\end{abstract}

\end{frontmatter}

%\linenumbers

\footnotetext[1]{Corresponding author: J. Michaud, michaud@cenbg.in2p3.fr}

\setcounter{footnote}{0}

\section{Introduction}

%\subsection{SPIRAL2 and DESIR}

DESIR \cite{DESIR} (Désintégration, Excitation et Stockage d’Ions Radioactifs) is the low energy facility under construction at GANIL (Grand Accélérateur National d’Ions Lourds) in France. It will allow to study the properties of exotic nuclei in unexplored regions of the nuclide chart. Exotic beams will come directly from the SPIRAL1, SPIRAL2 or S$^3$ facilities. Beams with masses up to A = 240 and an energy of 60keV should be delivered to the HRS by the RFQ cooler SHIRaC \cite{Shirac} with a low energy dispersion ($< 1eV$) and a transverse emittance of around $1\pi$ mm.mrad \cite{DESIR_TDR}. The High Resolution Separator (HRS) located at the entrance of DESIR will perform a beam purification with a design mass resolution of $R = M_0 /\Delta M  = 20000$. 

%\subsection{State of the art of magnetic mass spectrometers}

High-purity beams are a key element to the success of many experiments in nuclear physics with radioactive beams. High-resolution magnetic mass separators have the advantage of being fast (flight-time through the instrument of a few micro-seconds) while providing a mass resolving power to separate at least nuclei with different mass numbers. These separators provide therefore a (first) purification of nuclear species. 

Some facilities are already taking advantage of high resolution mass separators like ISOLDE \cite{ISOLDE} at CERN (effective resolution of R = 6000) or the ISAC facility at TRIUMF \cite{TRIUMF} (effective resolution of R = 10000). The CARIBU \cite{CARIBU} installation at Argonne National Laboratory recently installed its HRS with a design resolution of $R = 22400$. 

Despite extensive efforts from the community, mass spectrometers have difficulties to reach very high resolution as it demands a fine and accurate correction of higher-order optical aberrations, which can be difficult to detect and evaluate. The DESIR HRS has been designed specifically towards aberration cancellation by its mirror symmetry and the presence of a central multipole.

\section{Lattice, diagnostics and instrumentation}

As a magnetic separator, the HRS design is mainly focused around the two magnetic dipoles responsible for the mass dispersion. The beam line is completed by a set of fully electrostatic optical elements that are independent of the mass, including a 48-pole electrostatic multipole for higher-order corrections. 
For operation and commissioning purposes, it is equipped with various diagnostics and instrumentation devices.

\begin{figure*}[!ht]
    \begin{center}
    \includegraphics[scale=0.61]{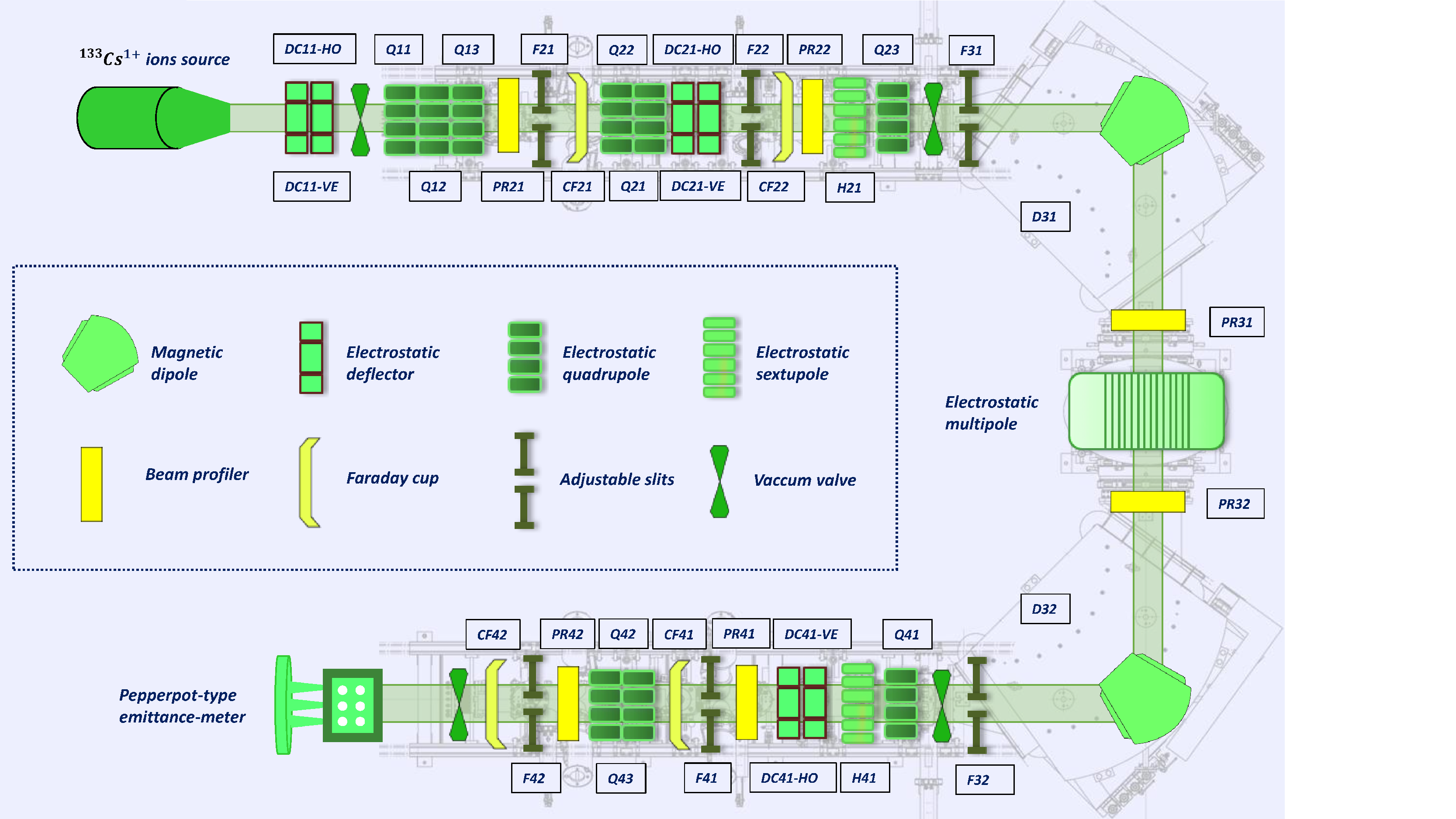}
    \caption{HRS synoptics as it is used for the commissioning. The picture is extracted from the HRS control system. See table \ref{25keVValues} for the values of an optimal tuning of the spectrometer with a 25 keV $^{133}Cs$ beam. For commissioning purposes, the entrance quadrupole triplet has been removed and replaced by a pair (vertical-horizontal) of steerers, named DC11. The ion source and the emittance-meter have been added to the synoptics as they are used for the commissioning.}
    \label{lattice}
    \end{center}
\end{figure*} 

All these HRS equipments are remotely controlled with an EPICS based Control System \cite{CCHRS}.
The HRS is now fully mounted at LP2I Bordeaux (former CENBG) and its commissioning started in July 2018.

\subsection{Lattice and elements}

The final optical design \cite{HRS} was validated in 2013 and consists of two 90$^o$ magnetic dipoles (D) with 36$^o$ ~entrance/exit angles, a set of matching quadrupoles (MQ), focusing quadrupoles (FQ), focusing sextupoles (FS) and a 48-pole multipole (M) in the symmetry plane of the beam line as shown in Fig. \ref{lattice}.

\textbf{The two dipoles} (D31 and D32) are normal conducting magnets with water-cooled coils and a 0.850 m curvature radius. They can analyse beams with masses up to A = 240 and energies up to 60keV.
The magnetic length as well as fringe fields can be controlled with a set of six correction coils (entrance, center and exit for both dipoles).

\textbf{The quadrupole doublets} (Q21/Q22 and Q42/Q43) are horizontally focusing (first quadrupole) - defocusing (second quadrupole) lenses, producing a narrow ribbon shaped beam with reduced Y angles, minimizing vertical angular aberrations.

The next \textbf{defocusing quadrupoles} (Q23 and Q41) tend to make the beam diverge horizontally so that the beam occupies the entire dipole magnet acceptance to maximize mass dispersion \cite{Wollnik}.

\begin{figure}[!ht]
    \begin{center}
    \includegraphics[scale=0.75]{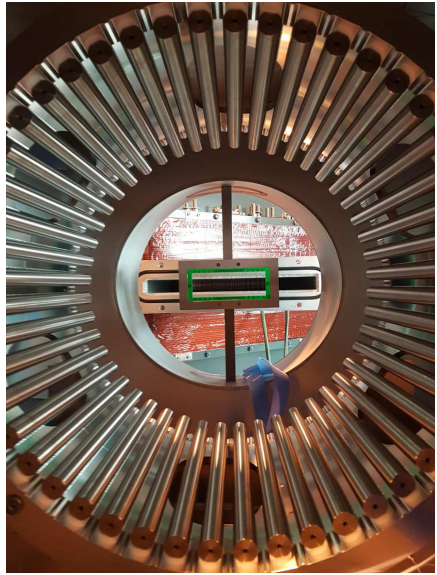}
    \caption{HRS multipole with its 48 electrodes. It has a length of 0.3m for an aperture of 0.4m. The intra-chamber profiler PR31 can be observed behind the multipole.}
    \label{multipole}
    \end{center}
\end{figure}

Finally, the HRS includes an \textbf{electrostatic multipole} (M) composed by 48 independent electrodes as shown in Fig. \ref{multipole}. 

Each electrode can be polarized independently, and the control system allows a superposition of multipolar fields up to the $5^{th}$ order with the formula of Eq. \ref{multipoleFormula}.

\begin{equation}
    V(P) = \sum_{n=2}^n V_n.\sin \left(n*\frac{2\pi(P+0.5)}{48}+\phi_n \right)
    \label{multipoleFormula}
\end{equation}

\noindent where $V(P)$ is the voltage applied on the pole $P\in [0 ; 47]$, $V_n$ the voltage of the $n^{th}$ order of the potential ($n^{th}+1$ order of the field) and $\phi_n$ is an optional phase. As examples, n = 2 is for a quadrupolar field, n = 3 for a hexapolar field etc. The 0.5 pole offset places the center of the horizontal virtual pole between the two electrodes closest to the horizontal plane.
Table \ref{25keVValues} contains the optimal tuning values of the HRS for a 25keV $^{133}Cs^{1+}$ ion beam.

\vspace{0.2cm}

\begin{center}
  \captionof{table}{HRS tuning values for a 25keV $^{133}Cs^{1+}$ ion beam considered as the standard optical setting. Electrostatic optical values for different energies can be immediately deduced by scaling the voltages with energy.\label{25keVValues}}
\begin{tabular}{|c|c|}
  \hline 
  \hspace{0.4cm} Electrostatic elements \hspace{0.4cm} & \hspace{0.4cm} Voltage (V) \hspace{0.4cm}  \\
  \hline
  \hline
  LHR-Q21 & -287.5  \\  \hline
  LHR-Q22 & 338.75  \\  \hline
  LHR-H21 & 0.0  \\ \hline
  LHR-Q23 & -371.25  \\  \hline
  LHR-M$_{hexa}$ & 160.0  \\  \hline
  LHR-M$_{octu}$ & 20.0  \\  \hline
  LHR-M$_{deca}$ & to be defined  \\  \hline
  LHR-M$_{dodeca}$ & to be defined  \\  \hline
  LHR-Q41 & -371.25  \\  \hline
  LHR-H41 & 0.0  \\  \hline
  LHR-Q42 & 338.75  \\  \hline
  LHR-Q43 & -287.5 \\  
  \hline
\end{tabular}
\end{center}

\vspace{0.2cm}

\begin{center}
\begin{tabular}{|c|c|c|}
  \hline 
  \hspace{0.2cm} Magnetic elements \hspace{0.2cm}&\hspace{0.2cm} Intensity (A)\hspace{0.2cm} & \hspace{0.2cm}Field (G)\hspace{0.2cm}  \\
  \hline
  \hline
  LHR-D31 & 110.141 &  3076.1 \\ \hline
  LHR-D32 & 109.940 &  3076.1 \\ \hline
\end{tabular}
\end{center}

\vspace{0.2cm}

%\textcolor{orange}{Suite à la remarque de Bertram, ce paragraphe peut-être enlevé. Je l'ai écrit car il est facile de faire le calcul du Brho pour le HRS et se rendre compte que ce n'est pas la bonne valeur de champ pour du Cs133 à 25keV, donc j'avais anticipé les questions.}
%The dipoles magnetic field required to transport the beam is different from the theoretical one (\textcolor{orange}{3085.6G}). This should be solved in the next future with the help of 6 correction coils (entrance, exit and inner part for both dipoles) that will allow correcting the magnetic length of the dipoles.

Additionally to the optical elements, other elements are used for the commissioning:

- \textbf{A $^{133}Cs^{1+}$ ion source} provides a mono-isotopic stable beam to the HRS up to 5keV with a low intrinsic energy dispersion ($< 1eV$). The beam energy can be increased up to 25keV with an additional high voltage platform. Its emittance $\epsilon_s$ was measured on a test bench for different energies and optical conditions. We estimated $\epsilon_s(3\sigma) \approx 10 \cdot \pi $.mm.mrad at 10keV, decreasing as the square root of the energy.
The source includes a pair of electrostatic steerers (horizontal and vertical) that allows us to center the position of the beam at the entrance of the HRS, but not to control its angle. 

To solve this issue, an additional \textbf{pair of steerers} (DC11) was mounted between the source and the HRS entrance. Hence, both the initial angle and the position can be controlled at the object point of the  spectrometer. The initial beam size can be controlled with an intra-source Einzel lens and with an extraction lens. 

\subsection{Diagnostics and instrumentation}

The HRS is equipped with sets of slits allowing a fine tuning of the beam :

\begin{itemize}
    \item Entrance slits (F21) located at the object point of the HRS are critical as they define the entrance emittance in the HRS. Closing these slits will reduce the object size, hence increase the spectrometer resolution, but decrease its transmission.
    \item An intermediate pair of slits (F22) located at the focal point between the quadrupole doublet and the defocusing quadrupole. They are useful to measure the very low beam size at this point and to align the beam before the dipoles.
    \item The dipole slits (F31) allow to control the beam size at the entrance of the first dipole.
    \item Slits F32 and F41 are respectively symmetric to F31 and F22.
    \item Exit slits (F42) are symmetric to F21 and allow the selection of the separated beam.
\end{itemize}

Additionally, the HRS is equipped with four Faraday Cups (CF21, CF22, CF41 and CF42) and four beam position monitors (PR21, PR22, PR41 and PR42), located at specific positions such as focal points. For commissioning purposes, two new beam profilers (PR31 and PR32) were recently added before and after the multipole to measure the beam size and the alignment within the two dipoles.

The final resolution of the HRS will mainly depend on the correction of optical aberrations. These aberrations can be studied through the beam emittance figure (see section \ref{SecAbb}), measured with an emittance-meter.
The emittance-meter that was chosen for the commissioning is a Pantechnik pepperpot-type emittance-meter \cite{PKemitt}. It is composed of a movable band of 20$\mu$m holes that can be used to measure the correlation between angles and positions of the beam.
The movable band is used to scan the horizontal positions of the beam with adjustable steps and avoid horizontal overlaps in the sampled data.

Data visualization uses a pair of micro-channel plates (MCP), a phosphor screen and a CCD camera.
Each pixel of the camera can be linked to a position on the MCP, and to an angle relative to the position of the pepperpot holes.
An emittance reconstruction and analysing software has been written specifically for the HRS commissioning and benchmarked with the Pantechnik analysis software.

\begin{figure*}[!ht]
    \begin{center}
    \includegraphics[scale=0.333]{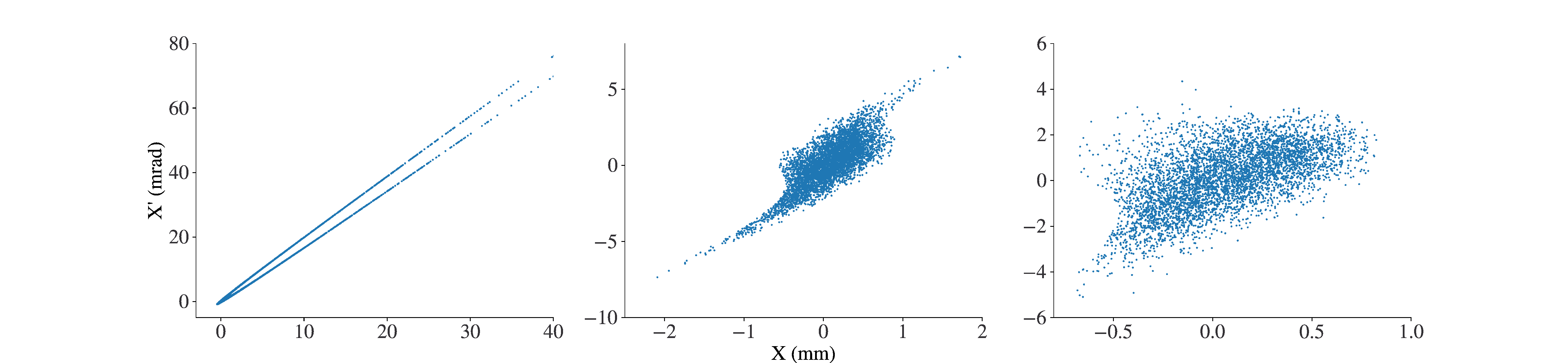}
    \caption{Simulated emittance figures at the image point of the HRS, for a 15keV $^{133}$Cs beam in the conditions of the optical design. From left to right: no correction applied (2nd order aberration dominant), hexapolar correction applied (3rd order aberration dominant), hexapolar + octupolar corrections applied (higher order aberrations dominant). }
    \label{emitTh}
    \end{center}
\end{figure*} 

\begin{figure*}[!ht]
    \begin{center}
    \includegraphics[scale=0.333]{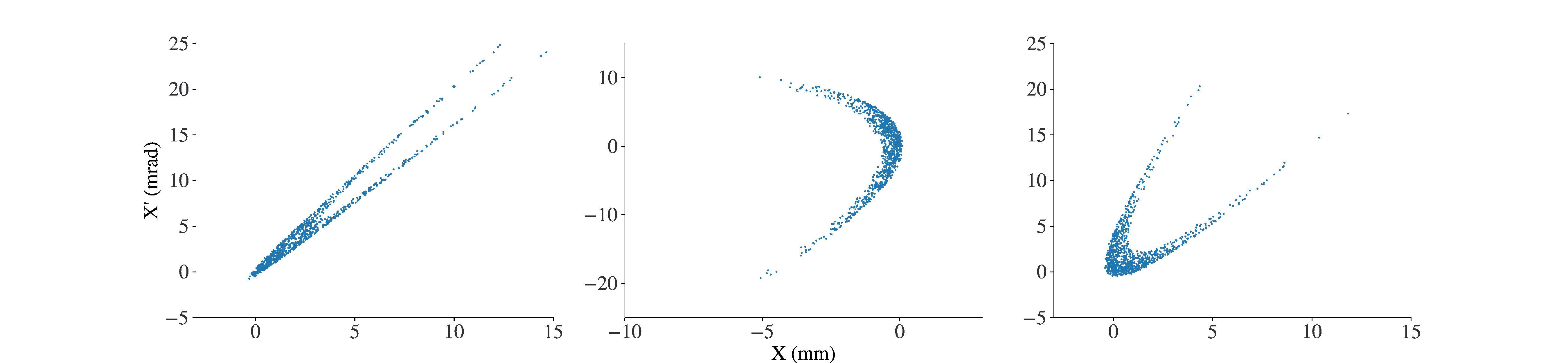}
    \caption{Comparison of the different measurement methods. Left: simulation of standard optical conditions with the emittance-meter placed after the image point of the HRS. Center: simulated emittance at the level of F41. Right: Q42 set to 100V and Q43 set to 0V to enlarge the "C" aperture of the hexapolar aberrations after the image point of the HRS.}
    \label{compaMethodsAb}
    \end{center}
\end{figure*} 

The HRS is also equipped with a set of RMN probes to monitor the nominal field of the dipoles.
By closing the final slits (F42), we can scan the beam through them with the dipoles and monitor the beam intensity over a Faraday cup (CF42) to have an extremely precise profile of the beam at the image point of the HRS. By this method, one ensures that the beam profile is exactly measured at the mass separation point of the HRS and with a fine sampling, as the dipole current supplies and RMN probes have a $10^{-6}$ accuracy. This method will mainly be used in the following to acquire beam profiles.

\section{High order aberrations and corrections}

\label{SecAbb}

Optical aberrations in the beam line will tend to increase the final beam size. In that case, the resolution given by Eq. \ref{theorique} is decreased \cite{Wollnik}: 

\begin{equation}
    R = \frac{D_\delta + D_\delta'L_2}{2x_{00}(D_x + D_x'L_2)} = \frac{D}{2x_{00}M + \delta}
    \label{theorique}
\end{equation}

\noindent where $D$ is the dispersion of the system, 2$x_{00}$ the entrance beam size, $M$ the magnification and $\delta$ stands for optical aberrations and systematic errors like misalignments or field inhomogeneities.

These aberrations have to be corrected up to the highest order in order to maximize the mass resolution. They can be measured by looking at the final horizontal beam size or on the phase space through emittance figures.

\subsection{Simulated and measured aberration figures}

The two-dimensional space of the emittance figure \{$x,x'$\}, where x is the particle position and x' its angle with respect to the longitudinal axis, shows the correlation between the positions and the angles of the particles of a beam.
Simulations with COSY infinity \cite{COSY} allow us to predict the shape of the emittance figure at any position in the beam line for different amounts of aberrations.
Fig. \ref{emitTh} shows the simulated emittance figures for a 25keV $^{133}Cs^{1+}$ beam with different types of aberrations and corrections, at the separation point of the HRS (slits F42).
For illustration purposes, it is generally assumed that a 2nd order aberration leads to a "C-shaped" emittance, and a 3rd order aberration to a "S-shaped" emittance.

The intrinsic resolution of the emittance-meter is not sufficient to separate the tails of the second order aberrations at the image point of the HRS under the conditions of the standard optical setting. Two solutions, illustrated by Fig. \ref{compaMethodsAb}, have been envisaged:

\begin{itemize}
    \item Move the emittance-meter to another position in the beam line, where the aberration figures are more explicit. Simulations show maximum apertures for the "C-shaped" emittance at the center of the multipole and at the focal point of the second focusing quadrupole Q41, where the slits F41 are placed. As the second order aberrations mainly come from the dipoles, the first position is not possible. Furthermore, mechanical integration does not allow yet to place the emittance-meter at the second position.
    \item Modify the focusing conditions of the second half of the HRS, i.e. after the second dipole, so that the quantity of aberrations does not change significantly. Simulations show that, in certain optical conditions, aberration tails of the emittance figures can be measured experimentally (by "opening" the aperture of the "C-shape") and are corrected by the same voltage of the multipole as in the standard setting. We define for the following the "asymmetric configuration" as the standard optics configuration of the HRS with Q42 set to 100V and Q43 set to 0V.
\end{itemize}

\begin{figure}[!ht]
    \begin{center}
    \includegraphics[scale=0.25]{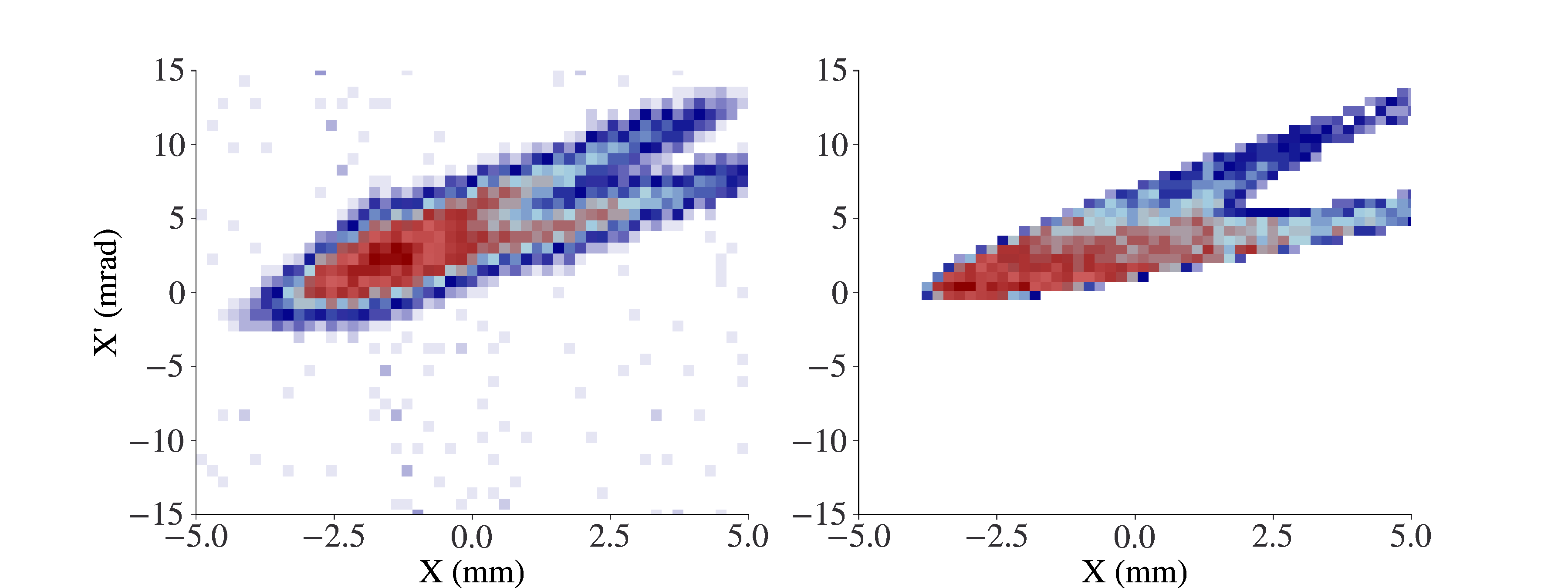}
    \caption{Measured (left) and simulated (right) emittance in asymmetric optical conditions of the HRS. For a 15keV beam, only the last quadrupole doublet has been changed (Q42 = 100V, Q43 = 0V).}
    \label{comparEm}
    \end{center}
\end{figure}

\noindent For the HRS commissioning, we chose to use the second method and change the optical conditions of the second half of the HRS. Fig. \ref{comparEm} shows an example of measured and simulated emittance figures in these asymmetric optical conditions.

We observe a good qualitative matching of experimental measurements with the simulations that was confirmed with different optical configurations and energies.

\subsection{2nd and 3rd order corrections}

\begin{figure}[!ht]
    \begin{center}
    \includegraphics[scale=0.58]{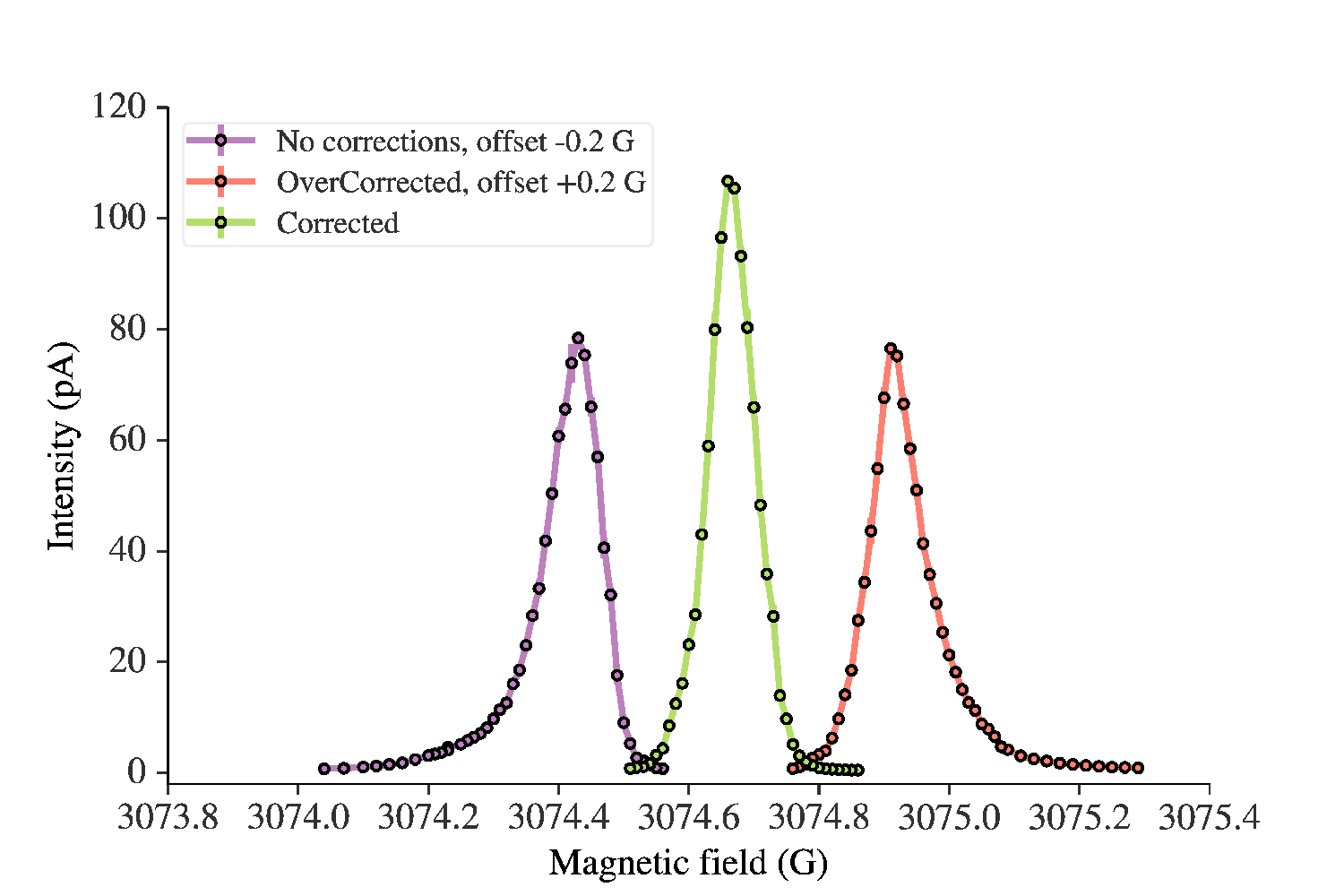}
    \caption{Beam position profiles acquired by scanning the beam with the dipoles through slits F42 and monitoring its intensity on CF42. With no correction (0V), the second order aberration tail extends toward smaller field values. With a correct level of correction (150V), the aberration tail disappears. If overly corrected (300V), the aberration tail extends towards larger field values. The first and the third curves are slightly displaced for better readability.}
    \label{fieldProfils}
    \end{center}
\end{figure}

\paragraph*{\textbf{2nd order: }} Second order aberrations are dominant compared to higher order aberrations. They create a parabolic ("C-shaped") aberration figure in the emittance phase space, whose aperture tends to decrease over the distance. This aperture can be increased artificially by a change of the optical conditions as mentioned above. A direct effect of this level of aberrations is an asymmetric increase of the beam size, hence a loss of resolution.
The use of a hexapolar electrostatic field in the center of symmetry of the HRS with the multipole can correct this type of aberrations. 
Fig. \ref{fieldProfils} shows the evolution of the beam position distribution (visualised here by a scan of the magnetic field as explained above) for different hexapolar fields of the multipole. With no corrections, the 2nd order aberration tail is visible for smaller field values (negative positions in the focal plane of the HRS), disappears with the correct level of hexapolar correction, and extends to higher field values (positive positions) when the second-order aberrations are overly corrected

\begin{figure*}[!ht]
    \begin{center}
    \includegraphics[scale=0.333]{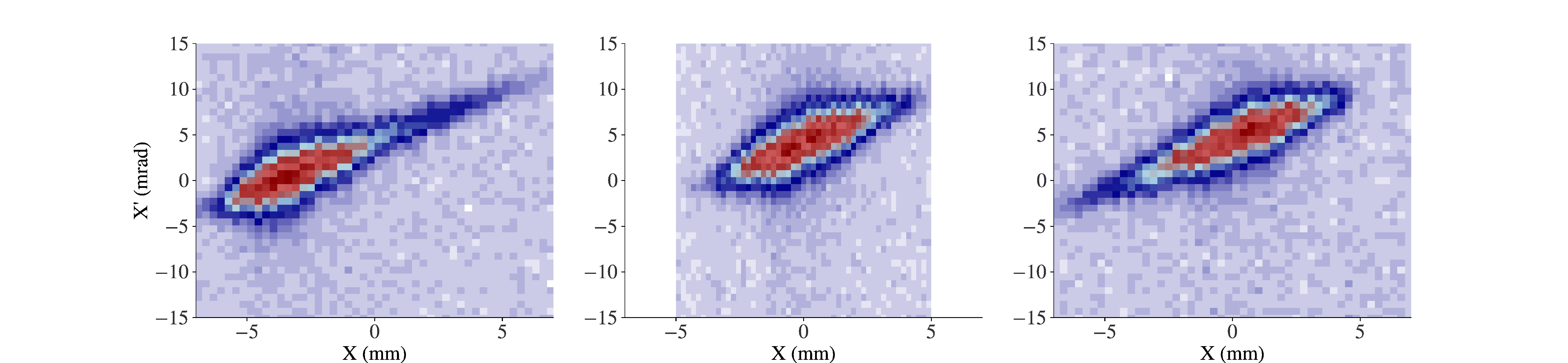}
    \caption{Measured emittance figures at the image point of the HRS for a 15keV $^{133}Cs$ beam in the standard optical design. From left to right: no correction applied, correct hexapolar correction applied, hexapolar term over-corrected.}
    \label{ComparaisonCorrectionsEmittances}
    \end{center}
\end{figure*} 

\begin{figure*}[!ht]
    \begin{center}
    \includegraphics[scale=0.333]{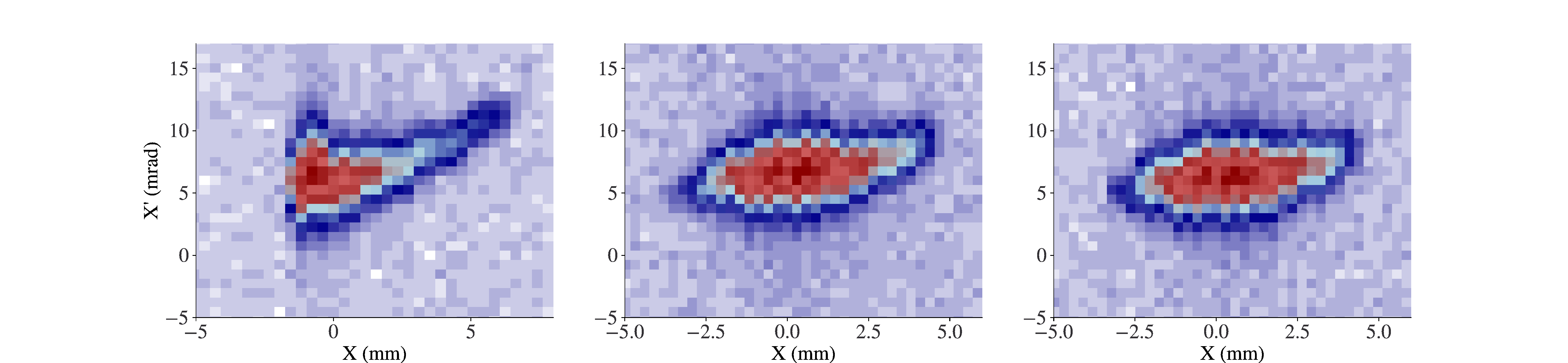}
    \caption{Measured emittance figures in the asymmetric configuration. The parabolic shape due to second order aberrations can be observed on the left picture. The central picture shows a third order tail after hexapolar corrections, and the right picture is corrected up to third order. }
    \label{emitt}
    \end{center}
\end{figure*} 

Fig. \ref{ComparaisonCorrectionsEmittances} shows the evolution of emittance figures with the same degrees of corrections in the standard optical configuration. The aperture of the aberration tail cannot be observed in this configuration, but the length of the aberration tail is still a good indicator of the second order aberration magnitude. The projection of the aberration tail on the position axis directly gives the beam profile distributions on the profiler PR42, which are similar to Fig. \ref{fieldProfils}.

\paragraph*{\textbf{Higher order aberrations }}

Fig. \ref{emitt} shows the evolution of emittance figures with different orders of corrections in the asymmetric configuration, where beam optics have been modified to be in the configuration shown in Fig. \ref{compaMethodsAb}. The second order "C-shape" can be observed as well as a bit of a third order "S-shape", but no mass separation can be achieved in this configuration. However, this mode is useful to apply corrections with the multipole.

\section{Current status and perspectives}

\subsection{Resolution measurements}

As the HRS source provides a mono-isotopic Cesium beam, a direct mass resolution of the HRS cannot be measured during the commissioning. 
However, the magnetic rigidity of the beam in the dipoles behaves similarly for mass or energy shifts and an energy difference can be used to measure the resolution. This can be seen in the magnetic rigidity formula $B \rho = p/q = \sqrt{2mE}/q$ or through the first order dispersion terms of the transfer matrix of the HRS where $<x:\Delta E/E> = <x:\Delta M/M> = -31 m$.

\begin{figure}[!ht]
    \begin{center}
    \includegraphics[scale=0.5]{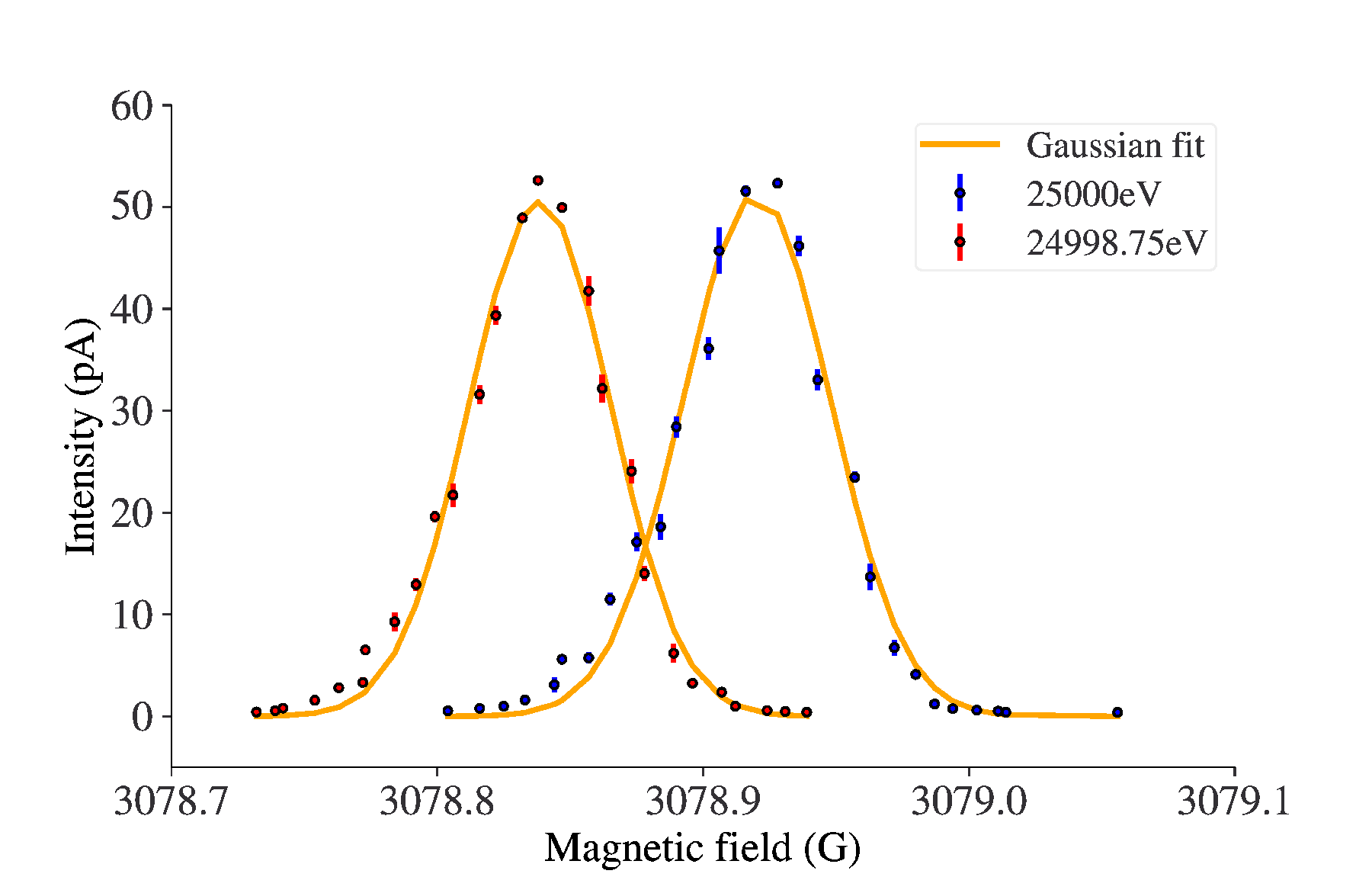}
    \caption{Resolution measurement for a 25keV Cesium beam. An energy shift of 1/20000 (1.25eV) was applied. Both beams are separated better than their FWHM. This is equivalent to a FWHM separation if the energy shift was 1/25000, leading to a FWHM resolution of R = 25000 for the HRS. }
    \label{Resolution25000}
    \end{center}
\end{figure}

As the outcome of a resolution measurement depends on many factors, we will list in the following the experimental conditions of the present measurement results:

\begin{itemize}
    \item The energy dispersion of the beam has been reduced to its minimum. The ion source has an intrinsic energy spread $< 1eV$ and the acceleration high voltage supply has been stabilized to $10^{-3}V$.
    \item The beam emittance has been measured on a test bench and is estimated to be around 1-2 $\pi.mm.mrad$ at 25keV. 
    \item Entrance slits (FH21/FV21) have been set to $\pm 0.5$mm both for the horizontal and the vertical axes while exit slits have been closed to $\pm 0.2$mm to guarantee a fine sampling for the beam profile acquisitions (field and slits method).
    \item The beam transmission from entrance to exit of the HRS has been estimated to be around $70\%$ when both entrance and exit slits are set to $\pm 0.5$mm.
    \item The resolution measurements compare two beams of same intensity, and estimates the mass/energy shift corresponding to a FWHM separation of these two beams.
\end{itemize}
\ \\ We used simultaneously three different methods to measure the resolution:\paragraph*{Magnetic field scan with 2 beams :} By slowly ramping both dipoles with fine steps, the beam will pass through the exit slits FH42 of the HRS. By monitoring the beam current on FC42 behind the slits, we obtain a precise beam profile with adjustable steps. This requires a fine current-field calibration of the dipoles to ensure both fields stay equal at any moment and a constant monitoring of the magnetic field with the RMN probes. 
The resolution measurement consists in repeating this measurement for a different energy and compare the beam profiles to extract the resolution. Fig. \ref{Resolution25000} shows such a resolution measurement for a Cs beam with an applied $\Delta E /E = 1/20000$.

%A cross-product of the beam FWHM and beam separation gives a FWHM resolution of R = 25000.
A Gaussian fit of the beam profiles gives a separation of $\Delta \mu = 0.082G$ with a standard deviation of $\sigma = 0.027G$. This is equivalent to a FWHM separation if the energy shift was 1/25000, leading to a FWHM resolution of $R = 25000$ or a 10\% valley resolution of $R =  13666$ for an entrance beam of 1 mm and a 1-2 $\pi.mm.mrad$ emittance.

\paragraph*{Magnetic field scan with one measurement:} The resolution corresponding to a beam separation at N times the standard deviation ($N \sigma$) can also be defined as: 

\begin{equation}
    R = \frac{1}{2} \cdot \frac{N \sigma_B}{\mu_B}
\end{equation}

\noindent where $\sigma_B$ is the standard deviation of the distribution of the beam intensity as a function of the field scan and $\mu_B$ its mean value. This formula as well as the factor $\frac{1}{2}$ directly comes from the derivation of the $B \rho$ formula. 
Applied to the $25keV$ measurement of figure \ref{Resolution25000}, with $\sigma_B = 0.027G$  and $\mu_B = 3078.9G$  we obtain a FWHM resolution of $R = 24200$ for an entrance beam of 1 mm and a 1-2 $\pi.mm.mrad$ emittance in agreement with the first method.

\begin{figure}[!ht]
    \begin{center}
    \includegraphics[scale=0.58]{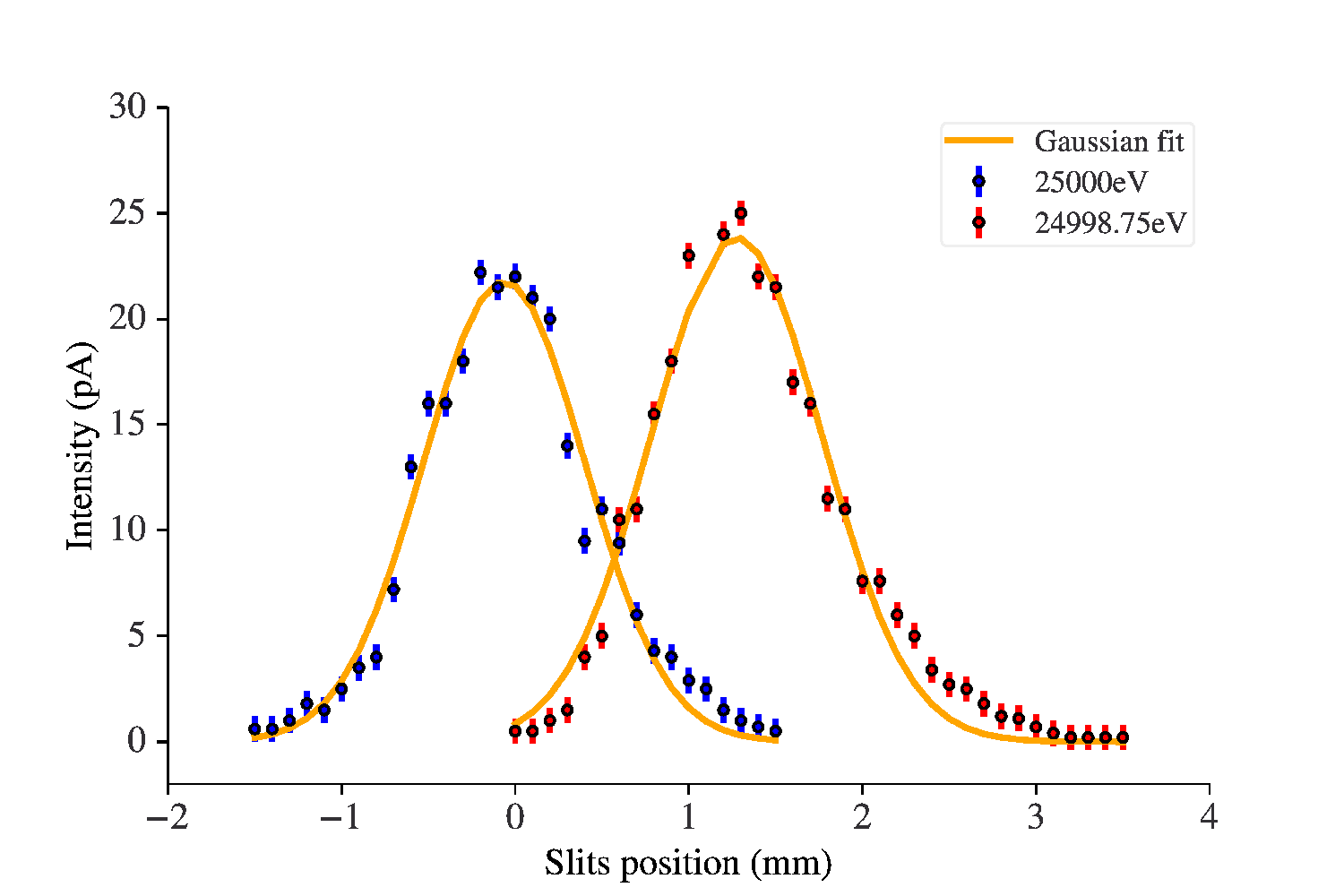}
    \caption{Resolution measurement for a 25keV Cesium beam. The beam positions have been sampled with the exit slits of the HRS (FH42) and an energy shift of 1/20000 (1.25eV) was applied between the two measurements. Both beams are separated better than their FWHM and this measurement leads to a FWHM resolution of R = 24000.}
    \label{ResolutionSlits}
    \end{center}
\end{figure}

\paragraph*{Beam position profiles with different energies:} a resolution measurement can be achieved by measuring the position profile of two beams with close energies. As the profiler PR42 is not at the focal point of the HRS, it cannot be used to do such a measurement. However, the slits FH42 can be used to scan the beam intensity on CF42 as a function of their position. This method is closer to a real mass separation as it does not require to vary the magnetic field of the dipoles but it is limited by the slits motorization. Fig. \ref{ResolutionSlits} shows a resolution measurement that was done with this method with two consecutive measurements at 25000eV and 24998.75eV (energy shift of $\Delta E / E = 1/20000$).

With a beam separation of $\Delta \mu = 1.33mm$ and a $FWHM = 1.08mm$ we deduce a FWHM resolution of R = 24000 for this measurement.

\subsection{Development program}

\begin{figure*}[!ht]
    \begin{center}
    \includegraphics[scale=0.319]{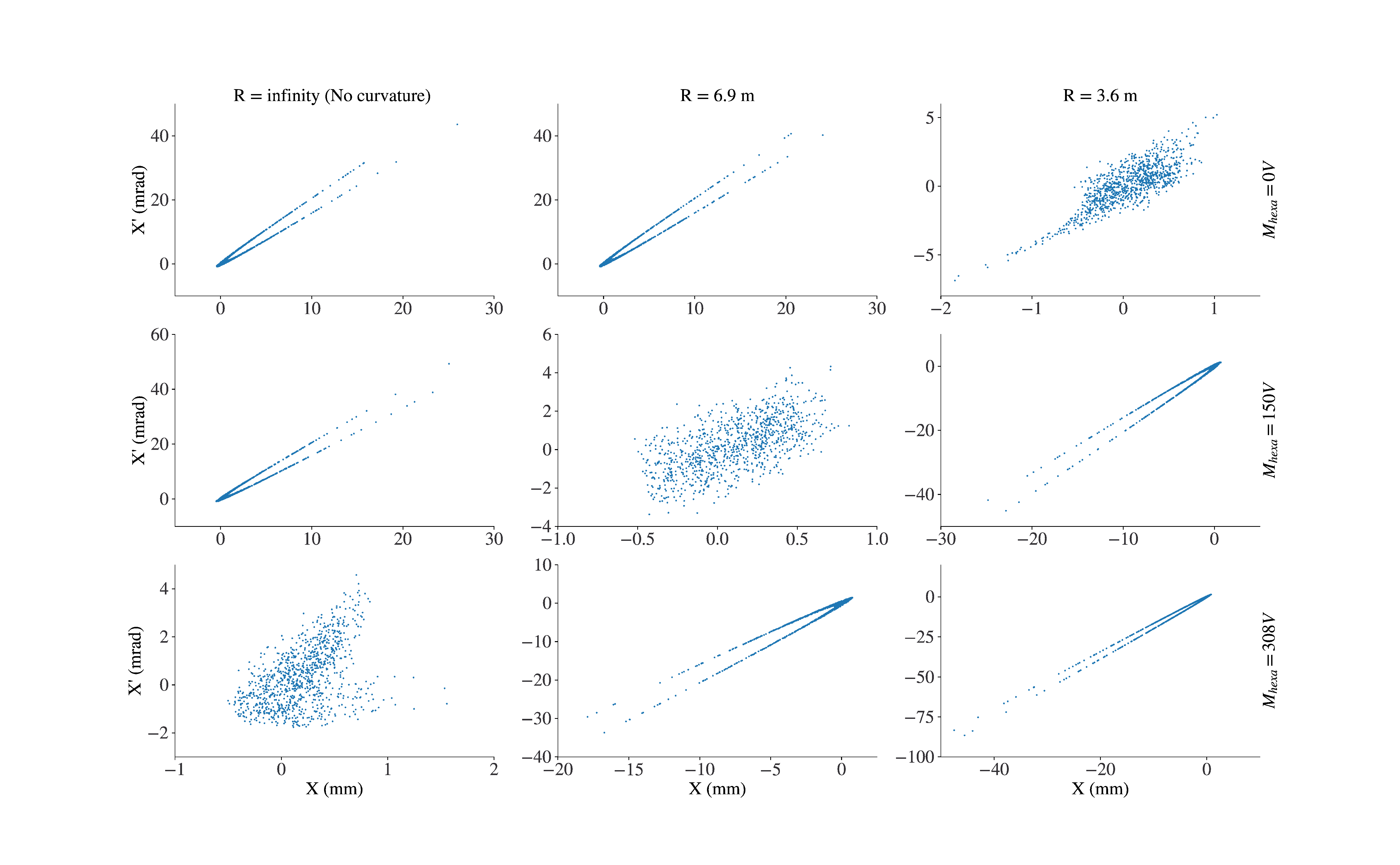}
    \caption{Emittance figures for different multipole hexapolar voltages and different poles curvatures. The left column corresponds to an infinite curvature radius (flat poles), the center column to a 6.9m curvature radius and the right column to a 3.6m curvature radius. The upper line corresponds to a 0V hexapolar correction on the multiple, the center line to a 150V correction and the bottom line to a 308V correction. A curvature radius of 6.9m has been chosen for the inner poles to compensate the experimental 150V multipole correction.}
    \label{Courbures}
    \end{center}
\end{figure*}

\paragraph*{\textbf{Pole reshaping:}} The 36$^o$ entrance/exit poles of the dipoles are flat poles and can be removed and reshaped to a more circular shape to naturally correct second order aberrations introduced by the dipoles. This will have three main advantages : 

\begin{itemize}
    \item It will simplify the tuning of the machine as a significant part of the hexapolar aberrations will already be corrected. Minor corrections with the multipole would probably still be necessary.
    \item The corrections will be permanent and will not necessitate a rescaling with energy.
    \item Applying strong hexapolar electric fields with the multipole, as done presently, implies the appearance of higher order aberrations. A reshaping of the poles should limit the amount of higher order corrections necessary for the HRS tuning.
\end{itemize}
Simulations give an optimal hexapolar correction on the multipole for a 300V hexapolar potential, while it is around 150V experimentally. It is most likely due to a slight geometrical deformation of the dipole iron yoke compared to the calculated one. Hence, we choose to fit with simulations for an optimal radius of the poles with a +150V hexapolar voltage applied on the multipole, leading to a 150V-equivalent remaining correction. For technical reason, and because the inner part of the dipoles are more sensitive to pole reshaping \cite{HRS} we decided to rework only the inner poles of the dipoles (multipole side).
Simulations with Zgoubi \cite{zgoubi} give an optimal correction radius of 6.94 m for the inner poles of the dipoles. These values have been confirmed with COSY simulations as shown in Fig. \ref{Courbures}.

\paragraph*{\textbf{CorrAb: Emittance figure-based auto-tuning of the multipole}}

Emittance figures are one of the best indicators for higher-order aberrations in a beam, and their analysis requires specific tools.
CorrAb is an analysing and auto-tuning software under development at LP2IB. It is planned to recognize patterns such as the so-called "C-shape" or "S-shape" or even higher orders and associate them with correction voltages on the multipole. By repeating emittance measurements, analysis, and corrections with the multipole, the program should be able to converge to a corrected tuning of the HRS.

\paragraph*{\textbf{Magnetic length correction}}

Six additional correction coils of the dipoles (entrance, center and exit for each dipole) will be used in order to correct the fringe field profiles, magnetic length and the nominal field in the center of the dipoles. At 25keV, the field necessary to pass the beam through the HRS is different from the theoretical field by around 9G and should be corrected by the correction coils in the near future.

\section*{Conclusion}

The DESIR HRS has been developed based on the feedback of other high-resolution mass separators like the CARIBU HRS at ANL. Its layout is mirror-symmetric around an electrostatic multipole, allowing a fine control of optical aberrations. Special care was taken with respect to the dipole design (especially concerning the field homogeneity and misalignments control). 

Since the beginning of its commissioning in 2018, we managed to understand and correct optical aberrations up to $3^{rd}$ order, mainly through emittance measurements and simulations. Thanks to the many diagnostics of the beamline (six profilers, four Faraday Cups, four RMN probes, six sets of slits and an emittance-meter) the beam in the HRS can be controlled with precision from the object to the image point and provide an optimal mass separation. 

Up to now, a FWHM energy resolution of R = 25000 with a transmission of about 70\% has been achieved for a 1mm $^{133}Cs$ beam at an energy of 25 keV with a very low energy dispersion, which demonstrates that the DESIR HRS is one of the best in-flight separators currently in operation.


\begin{thebibliography}{10}

\bibitem{DESIR}
DESIR website, available at http://www.cenbg.in2p3.fr/desir/, (accessed
  on March 19th, 2022).

\bibitem{Shirac}
R.~Boussaid, G.~Ban, J.~F. Cam, Experimental study of a high intensity
  radio-frequency cooler, Phys. Rev. Accel. Beams 18 (2015) 072802.
  doi:10.1103/PhysRevSTAB.18.072802.

\bibitem{DESIR_TDR}
B.~Blank, J.-C. Thomas, Technical proposal for SPIRAL2 instrumentation, DESIR:
  the SPIRAL2 low-energy beam facility. https://heberge.lp2ib.in2p3.fr/desir/IMG/pdf/DESIR-Technical-Proposal-V090105.pdf

\bibitem{ISOLDE}
R.~Catherall, et~al., The {ISOLDE} facility, Journal of Physics G: Nuclear and
  Particle Physics 44 (2017) 094002.
  doi:10.1088/1361-6471/aa7eba.

\bibitem{TRIUMF}
R.~Baartman, J.~Doornbos, Mass separator for the ISAC project at TRIUMF, 
  in: Proc. 5th European Particle Accelerator Conf. (EPAC96), CRC
  Press, Sitges, Spain, 1996.

\bibitem{CARIBU}
C.~Davids, D.~Peterson, A compact high-resolution isobar separator for the
  CARIBU project, Nucl. Instrum. Meth. B 266 (2008) 4449.
  doi:10.1016/j.nimb.2008.05.148.

\bibitem{CCHRS}
L.~Daudin, P.~Alfaurt, A.~Balana, et~al., CENBG control system and specific
  instrumentation developments for SPIRAL2-DESIR setups, in: Proc. 18th Int.
  Conf. on Accelerator and large Exp. Phys. Control Systems (ICALEPCS 2021).

\bibitem{HRS}
T.~Kurtukian-Nieto, et al., 
  %R.~Baartman, B.~Blank, T.~Chiron, C.~Davids, F.~Delalee,
  %M.~Duval, S.~E. Abbeir, A.~Fournier, D.~Lunney, F.~M{\'e}ot, L.~Serani, M.-H.
  %Stodel, F.~Varenne, H.~Weick, 
  SPIRAL2/DESIR high resolution mass separator,
  Nucl. Instrum. Meth. B 317 (2013) 284.
  doi:10.1016/j.nimb.2013.07.066.

\bibitem{Wollnik}
H.~Wollnik, A "Q-value" for particle spectrometers, Nucl. Instrum. 
  Meth. 95 (1971) 453.
  doi:10.1016/0029-554X(71)90545-3.

\bibitem{PKemitt}
H.~Kremers, J.~Beijers, S.~Brandenburg, A versatile emittance meter and profile
  monitor, in: Proceedings of the 8th European DIPAC07, Venice, Italy, 2008.

\bibitem{COSY}
M.~Berz, K.~Makino, COSY INFINITY version 8.1,  User's guide and reference manual
  (04/2001).

\bibitem{zgoubi}
F.~M\'eot, "ZGOUBI user's guide", C-AD/AP/470, Brookhaven National
  Laboratory, Upton (2013).

\end{thebibliography}
\end{document}